\documentclass{article}
\usepackage[
top    = 2.7cm,
bottom = 3.0cm,
left   = 2.5cm,
right  = 2.5cm]{geometry}
\usepackage{graphicx}
\graphicspath{%
    {converted_graphics/}
    {C:/Dropbox/1-kgl-top/Papers/2015/SurfaceDiffusionLength/}
}
\begin{document}

\begin{center}
{\LARGE The Surface Diffusion Length of Water Molecules on Faceted Ice:}%
\vskip6pt

{\LARGE A Reanalysis of \textquotedblleft Roles of Surface/Volume Diffusion}%
\vskip6pt

{\LARGE in the Growth Kinetics of Elementary Spiral Steps}\vskip5pt

{\LARGE on Ice Basal Faces Grown from Water Vapor,\textquotedblright }\vskip%
5pt

{\LARGE by Asakawa et al.}\vskip12pt

{\Large Kenneth G. Libbrecht}\vskip4pt

{\large Department of Physics, California Institute of Technology}\vskip1pt

{\large Pasadena, California 91125}\vskip-1pt

\vskip18pt

\hrule\vskip1pt \hrule\vskip14pt
\end{center}

\textbf{Abstract.} We reanalyzed the measurements made by Asakawa et al. 
\cite{furukawa14} of the growth velocities of single-molecule-high steps on
basal ice surfaces, as we believe the authors made a number of incorrect
assumptions regarding ice growth parameters and bulk diffusion in their
experiments. Applying what we believe are more accurate assumptions, we used
the data in \cite{furukawa14} to derive a surface diffusion length of $%
x_{s}\approx 10$ nm for water molecules on basal ice surfaces at $T=-8.4$ C,
about 500 times lower than what was reported in \cite{furukawa14}. Moreover,
in our analysis we found that no information about the height of the
Ehrlich-Schwoebel barrier could be obtained from these measurements.

\section{Introduction}

In \cite{furukawa14}, the authors describe a series of remarkable
measurements of the growth velocities of single-molecule-high steps on a
basal ice surface at $T=-8.4$ C. Specifically, they measured the step
velocity $v_{step}$ for a series of what can be approximated as equally
spaced steps, as a function of the spacing $L$ between steps. They also
measured $v_{step}$ as a function of the water vapor supersaturation $\sigma
_{ref}$ at a ice-coated reference surface that served as a water vapor
source. The experiments were conducted in air at normal atmospheric pressure.

In their analysis of these measurements, the authors found that the surface
diffusion length for water molecules on a faceted basal surface was $%
x_{s}\approx 5$ $\mu $m, and they found that the attachment coefficient for
water molecules on the basal surface near a step was $\alpha \approx 10^{-5}.
$ At low $\sigma _{ref},$ they measured the step kinetic coefficient
(defined by $v_{step}\approx \beta ^{L}\sigma _{ref})$ was $\beta
^{L}\approx 700$ $\mu $m/sec.

As described below, we believe that the authors substantially underestimated
the effects of bulk diffusion (of water molecules through the air) in the
analysis of their measurements. We also believe that the attachment
coefficient $\alpha \approx 10^{-5}$ they derived is strongly inconsistent
with a number of other measurements that reported $\alpha \approx 1$ for the
same quantity. We therefore reanalyzed the data in \cite{furukawa14} using
what we believe is an improved treatment of bulk diffusion, along with more
realistic assumptions regarding $\alpha .$

\begin{figure}[htb] 
  \centering
  \includegraphics[width=5.67in,height=2.59in,keepaspectratio]{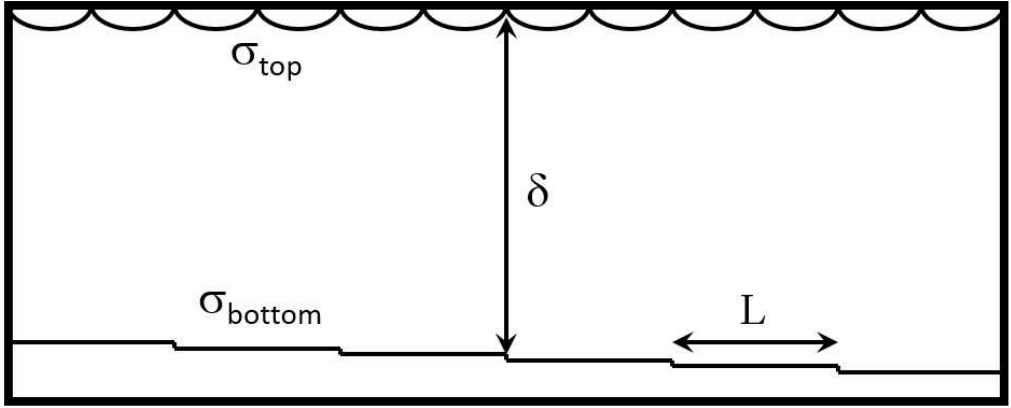}
  \caption{A schematic diagram of an
idealized experimental ice growth chamber. The top surface is covered with
small ice crystals at a temperature $T_{top},$ while the bottom surface at
temperature $T_{bottom}$ supports a faceted basal ice surface with a series
of molecular steps separated by a uniform spacing $L.$ The distance from top
to bottom is $\protect\delta ,$ and the supersaturation near the top surface
is a constant $\protect\sigma _{top}$ relative to the bottom ice surface.}
  \label{fig:Figure1}
\end{figure}

\section{Diffusion Analysis}

Consider the idealized experimental system shown in Figure 1, which is
similar to that described in the data analysis section in \cite{furukawa14}
(specifically the analysis described in Equations 1 through 11 in \cite%
{furukawa14}). We assume a constant supersaturation $\sigma _{top}$ at the
top surface of this simplified growth chamber (with the supersaturation
being measured relative to the temperature of the bottom surface), and we
assume that the bottom surface consists of a single faceted basal ice
surface containing a series of one-molecule-high steps with spacing $L.$ As
in \cite{furukawa14}, the step velocity can be written 
\begin{equation}
v_{step}=\frac{L}{a}\left\langle v_{basal}\right\rangle 
\end{equation}%
where $a$ is the size of a water molecule and $\left\langle
v_{basal}\right\rangle $ is the average perpendicular growth velocity of the
faceted surface.

Using a slightly different notation from \cite{furukawa14} (defined in
detail in \cite{libbrechtreview05}), solving the diffusion equation in this
simplified one-dimensional geometry gives%
\begin{equation}
\frac{c_{sat}}{c_{ice}}D\left[ \frac{\sigma _{top}-\sigma _{bottom}}{\delta }%
\right] =\left\langle v_{basal}\right\rangle =\frac{a}{L}v_{step}
\label{diff1}
\end{equation}%
which is equivalent to Equation 5 in \cite{furukawa14}. As in \cite%
{furukawa14}, we estimate $\delta \approx 200$ $\mu $m as a reasonable model
of the idealized system, as this is roughly equal to the size of, and
spacing between, the numerous test crystals on the sample surface.

We simplify the analysis slightly relative to \cite{furukawa14} by writing $%
v_{step}$ as \cite{libbrechtreview05}%
\begin{equation}
v_{step}\approx \frac{x_{s}}{a}\alpha v_{kin}\sigma _{bottom}  \label{diff2}
\end{equation}%
where $x_{s}$ is the diffusion length and $\alpha $ is the attachment
coefficient for water molecules striking the ice surface near the step (on
the lower terrace), specifically at distances small compared to $x_{s}.$
This is essentially equivalent to Equation 7 in \cite{furukawa14}, but in
the limit $x_{s}/L\ll 1.$

At this point our analysis begins to differ substantially from that in \cite%
{furukawa14}. First, we expect that $\sigma _{top}$ at the top of the
idealized box in Figure 1 is much smaller than the supersaturation $\sigma
_{ref}$ calculated at the distant vapor reservoir in the experiment. (In 
\cite{furukawa14}, $\sigma _{ref}=(P_{H_{2}O}^{\infty }-P_{e})/P_{e}$).
Second, we expect that $\alpha \approx 1,$ compared to the value $\alpha
\approx 10^{-5}$ reported in Figure 10 in \cite{furukawa14}. Needless to
say, these are large differences in our assumptions pertaining to the same
experimental data, which leads us to much different conclusions regarding
the surface diffusion length. We proceed by examining these assumptions in
more detail.

\subsection{The near-surface supersaturation}

Consider first an estimate of $\sigma _{top}.$ Our idealized box is quite
small, with $\delta =200$ $\mu $m, so $\sigma _{top}$ really represents the
supersaturation quite near the test crystals. As described in the Supporting
Information associated with \cite{furukawa14}, the sample surface is several
millimeters in size and is covered with a large number of growing ice
crystals, some of which are unobserved, and many of the ice crystals on the
sample surface are likely not completely faceted. In addition, the ice
crystals making up the vapor reservoir are about 16 mm away from the test
crystals, in a test chamber with a somewhat complex geometry. The vapor
pressure $P_{H_{2}O}^{\infty }$ defined in \cite{furukawa14} is the vapor
pressure at the vapor reservoir, which is not the same as the vapor pressure 
$\sigma _{top}$ just above the sample surface. The large number of crystals
on the sample surface all act as sinks, reducing the vapor pressure relative
to $P_{H_{2}O}^{\infty }.$ As stated in \cite{furukawa14}, \textquotedblleft
the existence of many crystals and a small amount of nonfaceted faces made
precise analysis of a volume diffusion field of water vapor impossibly
difficult.\textquotedblright\ We agree with this statement, and it means
that the value of $\sigma _{top}$ (essentially the value of $\sigma $ at a
height $\delta $ above a single faceted test crystal surface) is difficult
to determine. We would add that the geometry of the test chamber, together
with the large number of crystals on the sample surface, suggests that $%
\sigma _{top}$ is much smaller than $\sigma _{ref}$.

\begin{figure}[tbh] 
  \centering
  \includegraphics[width=5.5in,height=3.82in,keepaspectratio]{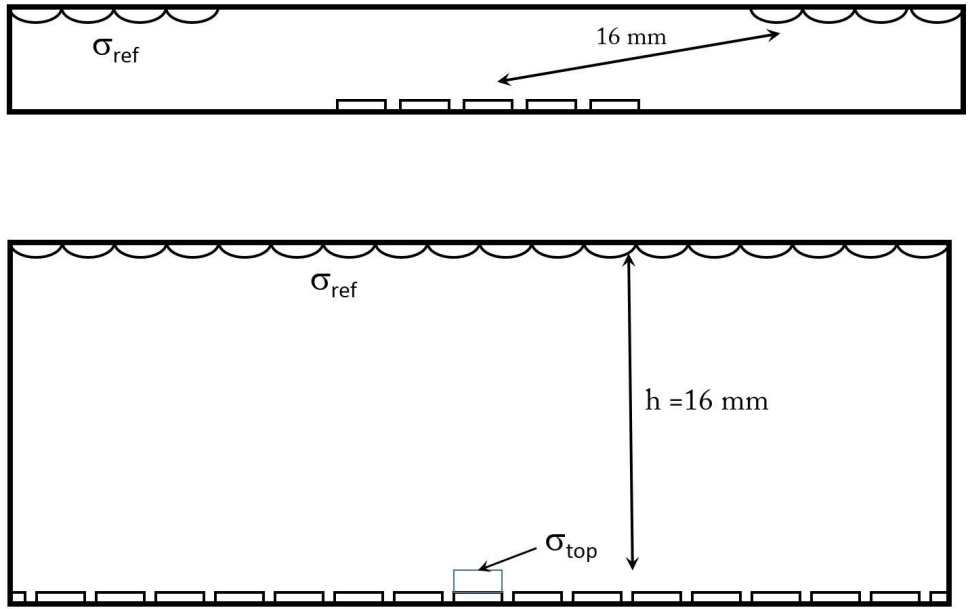}
  \caption{The upper sketch shows a crude
schematic approximation of the actual experimental chamber described in
detail in the Supporting Information associated with \protect\cite%
{furukawa14}. The sample surface is quite large and contains a large number
of ice crystals. The lower sketch is our idealized version of the same
chamber, simplified for ease of calculation. Within this large chamber we
embed the smaller test chamber in Figure 1, which has $\protect\sigma _{top}$
on the top surface and a single basal facet on the bottom surface.}
  \label{fig:Figure2}
\end{figure}

We can examine the bulk diffusion problem further by embedding the tiny
idealized test chamber in Figure 1 inside a much larger chamber shown in
Figure 2, which better represents the actual experimental chamber geometry
described in detail in the Supporting Information associated with \cite%
{furukawa14}. For $h$ we assume a value of 16 mm from \cite{furukawa14}, so
now we see that $\sigma _{top}$ is essentially the supersaturation just
above a large field of growing test crystals. Performing a diffusion
analysis similar to that above gives%
\begin{eqnarray}
\frac{c_{sat}}{c_{ice}}D\left[ \frac{G\sigma _{ref}-\sigma _{top}}{h}\right]
&=&\left\langle v_{basal}\right\rangle   \label{bigdiff} \\
&=&\left\langle \alpha _{bottom}\right\rangle v_{kin}\sigma _{top}  \nonumber
\end{eqnarray}%
where $G<1$ is a correction factor to account for the nontrivial geometry of
the test chamber compared to the more open geometry shown in the lower
sketch in Figure 2. Also $\left\langle \alpha _{bottom}\right\rangle $ is
defined by $\left\langle v_{basal}\right\rangle =\left\langle \alpha
_{bottom}\right\rangle v_{kin}\sigma _{top},$ so $\left\langle \alpha
_{bottom}\right\rangle $ is essentially the area-averaged attachment
coefficient of all the crystals on the sample surface, including nonfaceted
areas.

Rearranging Equation \ref{bigdiff} gives%
\begin{equation}
\sigma _{top}\approx \frac{G\alpha _{diff}}{\left\langle \alpha
_{bottom}\right\rangle +\alpha _{diff}}\sigma _{ref}  \label{alphadiff}
\end{equation}%
where 
\begin{eqnarray*}
\alpha _{diff} &=&\frac{X_{0}}{h}\approx 10^{-5} \\
X_{0} &=&\frac{Dc_{sat}}{v_{kin}c_{ice}}\approx 145\textrm{ nm}
\end{eqnarray*}%
In the most likely case that $\left\langle \alpha _{bottom}\right\rangle \gg
\alpha _{diff}$ (assuming there are a substantial number of nonfaceted ice
crystals on the sample surface, as the authors describe) Equation \ref%
{alphadiff} simplifies to%
\begin{equation}
\sigma _{top}\approx \frac{G\alpha _{diff}}{\left\langle \alpha
_{bottom}\right\rangle }\sigma _{ref}  \label{sigtop}
\end{equation}

Unfortunately, we have no good way to estimate $\left\langle \alpha
_{bottom}\right\rangle $. In ice-free regions and on perfectly faceted ice
surfaces, we expect $\alpha \approx 0,$ while nonfaceted regions would give $%
\alpha \approx 1.$ All growing ice surfaces must contain some molecular
steps, and each step contributes to increasing $\left\langle \alpha
_{bottom}\right\rangle .$ Assuming a fraction $10^{-4}$ of $\alpha \approx 1$
surfaces and $G\approx 0.1$ gives $\sigma _{top}\approx 10^{-2}\sigma _{ref}$%
, but this is just a very rough estimate. As stated in \cite{furukawa14},
determining $\sigma _{top}$ with greater accuracy is \textquotedblleft
impossibly difficult\textquotedblright\ without a better defined
experimental arrangement. However, our estimate that $\sigma _{top}$ is much
smaller than $\sigma _{ref}$ is certainly consistent with, and expected
from, the experimental details presented in \cite{furukawa14}.

\subsection{The attachment coefficient}

In \cite{furukawa14} the authors mention that their derived values of $%
\alpha \approx 10^{-5}$ are much smaller than the values of $\alpha \approx
0.15$ found in two references. We would add that the measurements in \cite%
{kglalphas13} are substantially improved over the older references, giving $%
\alpha \approx 1$ on basal surfaces in the absence of a nucleation barrier.
We note also that there is no nucleation barrier for molecules impinging
near a molecular step. We disagree with the statement in \cite{furukawa14}
that \textquotedblleft there is no value of $\alpha $ that can be directly
compared with ours.\textquotedblright\ The different experiments are all
measuring essentially the same $\alpha ,$ so they can all be compared, and
the preponderance of evidence supports $\alpha \approx 1$. Moreover, a value
of $\alpha \approx 10^{-5}$ in the absence of a nucleation barrier would
imply ice crystal growth velocities that are orders of magnitude below what
are commonly observed; the growth of an ordinary snowflake would take weeks
with such a low attachment coefficient!

\section{An Improved Model}

Assuming that the above conclusions are essentially correct, we begin with
the assumptions that $\sigma _{top}\ll \sigma _{ref}$ and $\alpha \approx 1,$
and from these reinterpret the measurements presented in \cite{furukawa14}.
Going back to the idealized growth chamber in Figure 1, we can rearrange
Equations \ref{diff1} and \ref{diff2} to obtain%
\begin{equation}
v_{step}=\frac{X_{0}x_{s}v_{kin}\sigma _{top}}{a\delta }\left[ \frac{1}{%
\left( x_{s}/L\right) +\left( X_{0}/\alpha \delta \right) }\right] 
\label{vstep}
\end{equation}%
Taking $L\rightarrow \infty $ gives the velocity $v_{isolated-step}$ of an
isolated step, and $v_{step}=v_{isolated-step}/2$ when%
\begin{equation}
\frac{x_{s}}{L_{1/2}}\approx \frac{X_{0}}{\alpha \delta }
\end{equation}%
Using the 329 Pa data in \cite{furukawa14}, we take $L_{1/2}\approx 10$ $\mu 
$m along with our assumptions of $\delta =200$ $\mu $m and $\alpha \approx 1$
to obtain $x_{s}\approx 10$ nm, 500 times smaller than the $x_{s}\approx 5$ $%
\mu $m obtained in \cite{furukawa14}. Moreover we see that $x_{s}$ depends
on $\delta ,$ so there is a substantial uncertainty in our derived $x_{s}$
that depends on the uncertainties in our solution of the bulk diffusion
equation in the experimental chamber. In a nutshell, the complex geometry of
the experimental chamber does not allow a very accurate measure of $x_{s},$
owing to bulk diffusion effects. We estimate that our value of $x_{s}\approx
10$ nm is perhaps only accurate to a factor of three.

Because we find $x_{s}\ll X_{0},$ this means that $\sigma $ at the ice
surface near the growing step is nearly equal to what we called $\sigma
_{bottom}$ above. In other words, bulk diffusion does not substantially
lower the supersaturation at the surface of 10-nm features, relative to the
supersaturation just above these features. This can be verified to a
reasonable approximation using the analytic solution of the diffusion
equation for an infinitely long growing cylinder \cite{kglca13}. This
verifies Equation \ref{diff2} above.

\subsection{A consistency check}

To many it seems counterintuitive that covering 0.1 percent of an area with $%
\alpha =1$ surface (0.1 percent because $x_{s}/L_{1/2}\approx 10^{-3})$,
while leaving the remaining 99.9 percent with $\alpha =0,$ would reduce the
supersaturation $\sigma _{bottom}$ by a factor of two. To see that this is
indeed reasonable, we again rearrange Equations \ref{diff1} and \ref{diff2}
to obtain

\begin{equation}
\sigma _{bottom}\approx \frac{\alpha _{diff}}{\left\langle \alpha
_{bottom}\right\rangle +\alpha _{diff}}\sigma _{top}
\end{equation}%
where 
\begin{equation}
\alpha _{diff}=\frac{X_{0}}{\delta }\approx 10^{-3}
\end{equation}%
In the limit of large $L$ (no molecular steps) we have $\left\langle \alpha
_{bottom}\right\rangle =0$ and $\sigma _{bottom}\approx \sigma _{top},$ as
expected. But if $\alpha \approx 1$ surfaces cover just a fraction $10^{-3}$
of the surface, then $\left\langle \alpha _{bottom}\right\rangle \approx
10^{-3}$ and $\sigma _{bottom}\approx \sigma _{top}/2.$ Thus even a quite
simple diffusion analysis supports our result of $x_{s}/L_{1/2}\approx
10^{-3}.$

\subsection{A second consistency check}

In the limit of large $L\ $in Equation \ref{vstep}, the velocity of an
isolated step becomes%
\begin{equation}
v_{isolated-step}\approx \frac{x_{s}}{a}\alpha v_{kin}\sigma _{top}
\end{equation}%
Again using the 329 Pa data in \cite{furukawa14}, we take $%
v_{isolated-step}\approx 12$ $\mu $m/sec along with $v_{kin}\approx 370$ $%
\mu $m/s and $\alpha \approx 1$ to obtain $\sigma _{top}\approx
10^{-3}\approx 10^{-2}\sigma _{ref},$ consistent with our rough estimate
above.

\subsection{A third consistency check}

We can also consider the case of very low $\sigma _{ref}.$ With slow growth
of the test crystals, one expects essentially all exposed surfaces to become
faceted, greatly reducing $\left\langle \alpha _{bottom}\right\rangle $ in
Equation \ref{alphadiff}, to the point that $\sigma \approx \sigma _{ref}$
throughout the growth chamber. In this case Equation \ref{diff2} becomes%
\begin{equation}
v_{step}\approx \frac{x_{s}}{a}\alpha v_{kin}\sigma _{ref}
\end{equation}%
so $\beta ^{L}=v_{step}/\sigma _{ref}$ (defined in \cite{furukawa14})
becomes 
\begin{eqnarray}
\beta ^{L} &\approx &\frac{x_{s}}{a}\alpha v_{kin} \\
&\approx &12000\textrm{ }\mu \textrm{m/sec}  \nonumber
\end{eqnarray}%
This is an upper limit, however, since $\beta ^{L}$ is smaller if $%
\left\langle \alpha _{bottom}\right\rangle $ is even slightly greater than
zero. Thus a small residual $\left\langle \alpha _{bottom}\right\rangle >0$
in the experiment could easily explain the measured $\beta ^{L}\approx 700$ $%
\mu $m/sec presented in \cite{furukawa14}.

The measured change in $v_{step}$ with $\sigma _{ref}$ at higher $\sigma
_{ref}$ (Figure 7 in \cite{furukawa14}) can be similarly understood by
considering how $\left\langle \alpha _{bottom}\right\rangle $ increases with 
$\sigma _{ref}.$ As $\sigma _{ref}$ increases from zero, the faceted
crystals begin to grow, so steps emerge on their surfaces, increasing $%
\left\langle \alpha _{bottom}\right\rangle $ and decreasing $\sigma _{top}$
following Equation \ref{sigtop}. Soon the ice growth becomes strongly
diffusion limited, resulting in non-flat surfaces with quite large $%
\left\langle \alpha _{bottom}\right\rangle $, thus reducing the slope $%
dv_{step}/d\sigma _{ref}$ to the degree shown in Figure 7 in \cite%
{furukawa14}.

\section{Conclusions}

We have reanalyzed the data presented in \cite{furukawa14}, and our analysis
yields a surface diffusion length $x_{s}\approx 10$ nm for water molecules
on a basal ice surface at $T=-8.4$ C. Our analysis assumed $\alpha \approx 1$
from the outset, a value that is indicated by several other ice growth
experiments, while we rejected the extraordinarily low value $\alpha \approx
10^{-5}$ reported in \cite{furukawa14}. Essentially all of the data
presented in \cite{furukawa14} can be explained in a reasonable and
self-consistent way using our model, as described above. Moreover, our model
does not distinguish whether admolecules are attaching to the step from the
upper or lower terrace, or both. Thus we find that the step velocity
measurements provide no useful information about the height of the
Ehrlich-Schwoebel barrier, contrary to what was concluded in \cite%
{furukawa14}. 

It has long been known, and was recently demonstrated with improved accuracy
in \cite{kglalphas13}, that growing small, isolated, faceted crystals in a
near-vacuum environment reduces the effects of bulk diffusion to much more
manageable levels, thus better revealing molecular kinetic effects.
Measuring the growth velocities of one-molecule-high steps in such an
experimental system would be a welcome next step toward understanding the
fundamental physics of ice growth dynamics.

\bibliography{C:/Dropbox/1-kgl-top/Papers/2015/SurfaceDiffusionLength\/ArXiv/kglbiblio3}
\bibliographystyle{unsrt}

\end{document}